\documentclass[conference]{IEEEtran}
\usepackage{cite}
\usepackage{verbatim}

  \usepackage[pdftex]{graphicx}
  \usepackage{caption}
  \usepackage{subcaption}
  \usepackage[font=footnotesize,labelfont=footnotesize]{subcaption}
  \usepackage[font=footnotesize,labelfont=footnotesize]{caption}
  \usepackage{epstopdf}

\usepackage[cmex10]{amsmath}
\usepackage{amssymb}
\usepackage{amsfonts}
\usepackage{relsize}
\usepackage[capitalise]{cleveref}
\usepackage{cleveref}

\usepackage{amsthm}

\Crefname{figure}{Fig.}{Figs.}

\usepackage{algpseudocode}

\usepackage{array}

\usepackage{mdwmath}
\usepackage{mdwtab}

\usepackage{stfloats}

\usepackage{amsfonts}
\usepackage[fleqn,tbtags]{mathtools}

\usepackage{multirow}

\def\dashfill{\cleaders\hbox{-~-}\hfill}

\usepackage[ruled]{algorithm2e}

\usepackage{color,setspace}
\definecolor{blue}{rgb}{0.2,0.3,0.8}

\def\dashfill{\cleaders\hbox{-~-}\hfill}
\makeatletter
\newcommand*{\rom}[1]{\expandafter\@slowromancap\romannumeral #1@}
\makeatother

\begin{document}
\title{Sensor Selection and Power Allocation via Maximizing Bayesian Fisher Information for Distributed Vector Estimation}

\author{\IEEEauthorblockN{Mojtaba Shirazi, Alireza Sani, and Azadeh Vosoughi}
\IEEEauthorblockA{Department of Electrical Engineering and Computer Science\\
University of Central Florida\\
Email: {\tt \small mojsh@knights.ucf.edu, Alireza@knights.ucf.edu, azadeh@ucf.edu}}}
\maketitle
\begin{abstract}
In this paper we study the problem of distributed estimation of a Gaussian vector with linear observation model in a wireless sensor network (WSN) consisting of $K$ sensors that transmit their modulated quantized observations over orthogonal erroneous wireless channels (subject to fading and noise) to a fusion center, which estimates the unknown vector. Due to limited network transmit power, only a subset of sensors can be active at each task period. Here, we formulate the problem of sensor selection and transmit power allocation that maximizes the trace of Bayesian Fisher Information Matrix (FIM) under network transmit power constraint, and propose three algorithms to solve it. Simulation results demonstarte the superiority of these algorithms compared to the algorithm that uniformly allocates power among all sensors.
\end{abstract}
%
\IEEEpeerreviewmaketitle
\section{Introduction} \label{Introduction}
Researchers have made tremendous contributions to the study of distributed estimation problem for energy constrained sensor networks \cite{Vosoughi_Sani_2016,Goldsmith_2006,AlRegib_2009,Giannakis_2008,Vandendorpe_2012}. Furthermore, significant progress has been made toward data selection approaches and specifically designing distributed sensor selection/activation algorithms to optimize the performance of resource constrained sensor networks \cite{Matin_2,Matin_3,Mahleqa_M3,IEEE_nights_saeid_comm_let,Boyd_sensor_selection_tsp2009,Leus_tsp_2015,Varshney_sensor_sel_and_colab_tsp_2015,Varshney_sensor_selection_tsp2016,Mitra_tsp_2008,IEEE_nights_Parsa2}. The authors in \cite{Boyd_sensor_selection_tsp2009,Leus_tsp_2015} studied the sensor selection problem that minimizes parameter estimation error, and described a heuristic algorithm, based on convex optimization, that approximately solves this problem. \cite{Leus_tsp_2015} focused on a general nonlinear observation model and formulated the sensor selection problem as the design of a sparse vector, considering several functions of the Cram\'{e}r-Rao Bound (CRB) performance measures. \cite{Varshney_sensor_sel_and_colab_tsp_2015} introduced a unified framework to jointly design the optimal sensor selection and collaboration schemes subject to a certain information or energy constraint. The authors in \cite{Varshney_sensor_selection_tsp2016,Mitra_tsp_2008} considered the problem of sensor selection for parameter estimation under a network transmit power constraint. In particular, \cite{Varshney_sensor_selection_tsp2016} considered a linear observation model with correlated measurement noises and sought optimal sensor activation algorithm by formulating an optimization problem, in which the trace of Fisher Information Matrix (FIM) is maximized subject to energy constraints. \cite{Mitra_tsp_2008} considered estimation of a function of a random parameter, and analyzed the effect of measurement noise variance on the optimal power allocation, for different network topologies. The authors in \cite{IEEE_nights_Parsa2} developed a multi-tier distributed computing infrastructure for sensor selection and data allocation using mobile device cloud, where all tasks are executed in parallel by sharing the workload among multiple nearby sensors.
%
\begin{figure*}[t]
	\centering
	\includegraphics[width=5.6in]{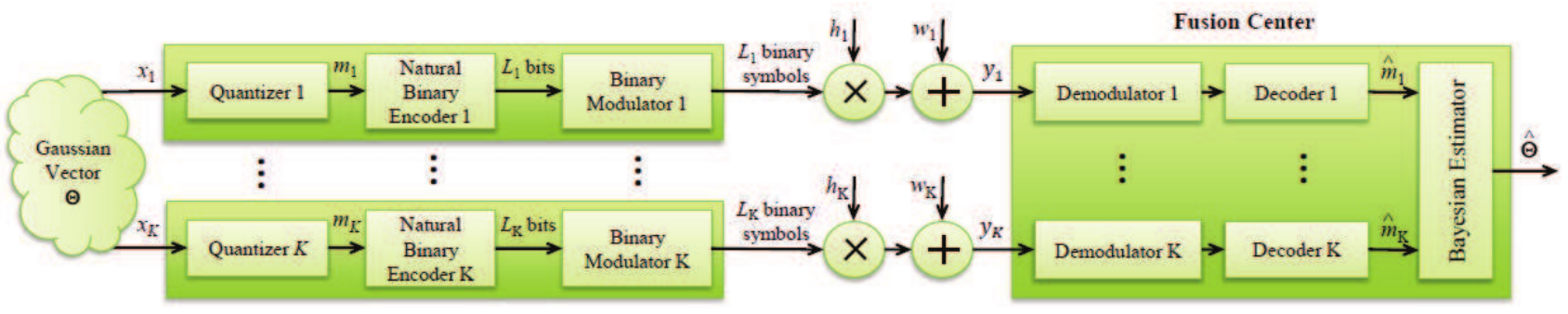}
	\vspace{-0.1cm}
	\caption{Our system model consists of $K$ sensors and a FC, that is tasked with estimating a Gaussian vector {\boldmath$\theta$}, via fusing collective received signals.}
	\label{system-model}
	\vspace{-.5cm}
\end{figure*}
%
In our previous works \cite{Shirazi_PIMRC2014}\cite{Shirazi_asilomar2014}, we presented our preliminary results on deriving Bayesian CRB matrix and studied the behavior of its trace, with respect to the system parameters. In our recent work \cite{Shirazi_Vosoughi_journal_2017}, considering the problem of distributed estimation of a Gaussian random vector with linear observation model, we derived the Bayesian FIM. We studied two transmit power optimization problems that maximize the trace of FIM and log-determinant of FIM under network transmit power constraint, and demonstrated that the estimation performance significantly enhances, using the transmit power allocation solutions corresponding to these two problems, compared with uniform power allocation among all sensors. Furthermore, a comparison was made between the estimation performance of coherent and noncoherent reception at the fusion center (FC).

{\it Adopting the same system model as in \cite{Shirazi_Vosoughi_journal_2017}, in this paper we study the sensor selection and power allocation problem that maximizes the trace of Bayesian FIM under network transmit power constraint.} We propose three algorithms to address this problem, and compare the performances of these algorithms in simulation results. 
\section{System Model and Problem Statement} \label{System Model}
We consider a network of $K$ sensors,
where each sensor observes a common zero-mean Gaussian vector {\boldmath$\theta$}= $[\theta_1, \theta_2,..., \theta_q]^T\!\in\!\mathbb{R}^q$ with covariance matrix $\boldsymbol{\mathcal C}_{\boldsymbol{\theta}}=\mathbb{E}\{\boldsymbol{\theta}\boldsymbol{\theta}^T\}$. 
We assume a linear observation model for sensor $k$ as:
\begin{equation} \label{obs_model}
x_k=\mathbf{a}_k^T \boldsymbol{\theta}+n_k, \ \ \ \ \ \ k=1,..., K
\end{equation}
where $\mathbf{a}_k\!=\![a_{k_1}, a_{k_2},..., a_{k_q}]^T\!\in\!\mathbb{R}^q$ is known observation gain vector and $n_k$ denotes zero-mean Gaussian observation noise with variance $\sigma_{n_k}^2$. We assume that $n_k$'s are uncorrelated observation noises across sensors and also are uncorrelated with {\boldmath$\theta$}. Sensor $k$ employs a uniform scalar quantizer with $M_k\!=\!2^{L_k}$ quantization levels $m_{k,l}\!=\!\frac{(2l-1-M_k)\Delta_k}{2}$ for $l\!=\!1,...,M_k$, where $\Delta_k$ denotes quantization step size and index $l$ indicates the quantization level $m_{k,l}$. We assume $p(|x_k|\!\geq\!\tau_k)\! \approx \! 0$ for some $\tau_k$ value. Hence, we choose ${\Delta_k} \!= \! \frac{2\tau_k}{(2^{L_k}-1)}$  \cite{Vandendorpe_2012,Vosoughi_Sani_2016}. The quantizer maps $x_k$ to one of the quantization levels $m_k \in \{m_{k,1},...,m_{k,M_k}\}$. 
%
%
Following quantization, sensor $k$ employs a fixed length encoder, which encodes the index $l$ corresponding to the quantization level $m_{k,l}$ to a binary sequence of length $L_k=\log_2 M_k$ according to natural binary encoding \cite{Vandendorpe_2012,Vosoughi_Sani_2016}, and finally modulates these $L_k$ bits into $L_k$ binary symbols. Let $P_k$ denote transmit power corresponding to $L_k$ symbols from sensor $k$, which is equally distributed among $L_k$ symbols. Each sensor employs a Binary Phase Shift Keying (BPSK) modulator, which maps each bit of $L_k$-bit sequence into one symbol with transmit power $P_k/L_k$.

Sensors send their modulated symbols to the FC over orthogonal flat fading channels, with complex-valued fading coefficient $h_k$. 
Suppose channel $h_k$ remains constant during the transmission of $L_k$ symbols. 
Denote $\nu_{k,i}$ as communication channel noise during the transmission of $i$-th symbol of $L_k$ symbols corresponding to sensor $k$. We assume $\nu_{k,i}$'s are independent and identically distributed across $L_k$ transmitted symbols and $K$ channels, $\nu_{k,i}\sim \mathcal{CN}\left(0,2\sigma_{\nu_k}^2\right)$. Furthermore, there is a constraint on the network transmit power, i.e., $\sum_{k=1}^{K}P_k\leq P_{tot}$, and thus, only a subset of sensors might be active at each task period.

Let $\hat{m}_{k}$ denote the recovered quantization level corresponding to sensor $k$, where in general, $\hat{m}_{k}\neq m_{k}$ due to communication channel errors. The FC processes channel output corresponding to sensor $k$ to recover transmitted quantization level $\hat{m}_k \in \{\hat{m}_{k,1},...,\hat{m}_{k,M_k}\}$. Having $\{\hat{m}_1,...,\hat{m}_K\}$, the FC applies a Bayesian estimator to form the estimate $\hat{\boldsymbol{\theta}}$. 
Under certain regularity conditions that are satidfied by Gaussian vectors, one can derive the $q\times q$ FIM, denoted here as $\boldsymbol{J}$. We refer the interested readers to section IV of \cite{Shirazi_Vosoughi_journal_2017} for the derivation of matrix $\boldsymbol{J}$. From (30) in \cite{Shirazi_Vosoughi_journal_2017}, we obtain:
\begin{eqnarray} \label{trace of J}
\text{tr}(\boldsymbol{J})&=&\text{tr}\left({\boldsymbol{\cal C}_{\boldsymbol{\theta}}}^{-1}\right)+\sum_{k=1}^{K}t_k(P_k),\\
t_k(P_k)&=&\frac{\mathbf{a}_k^T\mathbf{a}_k}{2\pi\sigma_{n_k}^2}\mathbb{E}_{\boldsymbol{\theta}}\{G_k(\boldsymbol{\theta},P_k)\},\nonumber\\
G_k(\boldsymbol{\theta},P_k)&=&\sum_{t=1}^{M_k}\frac{\left(\sum_{l=1}^{M_k}\alpha_{k,t,l}(P_k)\dot{\beta}_{k,l}
(\boldsymbol{\theta})\right)^2}{\sum_{l=1}^{M_k}\alpha_{k,t,l}(P_k)\beta_{k,l}(\boldsymbol{\theta})}.\nonumber 
\end{eqnarray}
Note that $t_k$ is a function of $P_k$. Due to the cap on the network transmit power,
only a subset of the sensors might be active at each task
period. So we introduce a sensor selection parameter $w_k\in\{0,1\}$, to indicate whether or not sensor $k$ is selected to participate in the distributed estimation task and transmit to the FC. {\it Our goal is to study sensor selection and transmit power allocation that maximizes tr($\boldsymbol{J}$), subject to network transmit power constraint. In other words, we are interested in solving the following constrained optimization problem}:
\begin{subequations}
\begin{align*} \tag{P1}\label{power aloc sen selection problem}
\mathop{\text{maximize}}_{P_k, w_k, \forall k}\ \ \ \ &\sum_{k=1}^{K}w_kt_k(P_k)\nonumber\\
\text{s.t.}\ \ \ \ &\sum_{k=1}^{K}P_k\leq P_{tot},\ P_k\in \mathbb{R}^{+},\ \forall k\\
&w_k\in \{0,1\},\ \forall k
\end{align*}
\end{subequations}
%
\section{Constrained Maximization of Bayesian Fisher Information} \label{Power Allocation} 
%
%
The constrained optimization problem in \eqref{power aloc sen selection problem} is a nonconvex mixed integer non-linear programming problem which is NP-hard combinatorial problem and its computational complexity
increases exponentially with the problem size. 

In the following, we propose three different algorithms to tackle \eqref{power aloc sen selection problem}. In the first two algorithms, we use two different relaxations of \eqref{power aloc sen selection problem} and in the third algorithm, we propose a reformulation of \eqref{power aloc sen selection problem} and solve it. As part of the first algorithm, relaxation \eqref{Boolean relaxed sensor selection} is obtained based on uniform transmit power allocation among all sensors and relaxation of the Boolean constraints in \eqref{power aloc sen selection problem}. Problem \eqref{Boolean relaxed sensor selection} becomes a linear programing problem and we solve it using simplex method. We refer to the first algorithm as uniform-select-uniform (USU) algorithm. In the second algorithm, we iteratively select a new sensor in a greedy manner until a stopping criteria is met. As part of the second algorithm, relaxation \eqref{max tr(J) for k in U_j} is obtained when we assume $\{w_k\}_{k=1}^{K}$ is given. Problem \eqref{max tr(J) for k in U_j} becomes a convex problem and we find the solution using Newton's method. We refer to the second algorithm as greedy algorithm. The third formulation \eqref{MCKP} is obtained based on discretizing transmit power and is called in the literature a multiple-choice knapsack problem (MCKP). Although \eqref{MCKP} is an NP-hard problem, we can find the solution in pseudo-polynomial time using dynamic programming. We refer to the third algorithm as MCKP algorithm. Define $\boldsymbol{t}\!=\![t_1,..., t_K]^T$, and let $\boldsymbol{P}\!=\![P_1,..., P_K]^T$ and $\boldsymbol{w}\!=\![w_1,..., w_K]^T$ be the vectors of sensors' powers and sensor selection parameters, respectively. Suppose $\boldsymbol{P}^{*}$ and $\boldsymbol{w}^{*}$ are final solutions for the algorithms.\\
$\bullet$ \underline{USU Algorithm}:
Initialize with $i=1$. Emplying uniform power allocation among all sensors, $\boldsymbol{P}$ and $\boldsymbol{t}$ can be obtained. Then, we find the best set of active sensors $S_{i+1}$ with cardinality $|S_{i+1}|=i$ via solving the following Boolean relaxed sensor selection problem:
\begin{align} \tag{P2}\label{Boolean relaxed sensor selection}
\mathop{\text{maximize}}_{\boldsymbol{w}}\ \ \ \ &\boldsymbol{w}^T\boldsymbol{t}\nonumber\\
\text{s.t.}\ \ \ \ &\boldsymbol{1}^T\boldsymbol{w}\leq i,\nonumber\\
&\boldsymbol{w}\in [0,1]^K.\nonumber
\end{align}
Given the set $S_{i+1}$ we uniformly allocate the power among the selected $i$ sensors, i.e., $P_k=P_{tot}/i,\ k\in S_{i+1}$ and repeat this procedure while incrementing $i$ until a stopping criteria is met. A summary is described in Algorithm \ref{USU algorithm}.\\
$\bullet$ \underline{Greedy Algorithm}:
Define ${\cal A}$ and ${\cal I}$ as sets of active and inactive sensors, respectively. Initialize with $i=1$, ${\cal A}\!=\!\{\}$, ${\cal I}\!=\!\{1, \dots, K\}$. For each sensor in the inactive set ${\cal I}$, say ${\cal I}_j,\ j=1, ..., |{\cal I}|$, use Newton's method to find the optimal solution $\{P^{'}_k\}_{k\in {\cal U}_j}$ for the following power allocation problem:
 \begin{align*} \tag{P3}\label{max tr(J) for k in U_j}
 \mathop{\text{maximize}}_{P_k,k\in {\cal U}_j}\ \ \ \ &\boldsymbol{1}^T\boldsymbol{t}\nonumber\\
 \text{s.t.}\ \ \ \ &\sum_{k\in {\cal U}_j}P_k\leq P_{tot},\ P_k\in \mathbb{R}^{+},\ \forall k\in {\cal U}_j,
 \end{align*}
where ${\cal U}_j\!=\!{\cal A}\cup {\cal I}_j$, and save $Y_{j}=\text{tr}\left(\boldsymbol{J}\left(\{P^{'}_k\}_{k\in {\cal U}_j}\right)\right)$. In \cite{Shirazi_Vosoughi_journal_2017}, we argue that \eqref{max tr(J) for k in U_j} is a convex programming problem. Then we select the sensor ${\cal I}_j$ which gives the largest performance improvement, i.e. the one corresponding to $max\{Y_{1}, ..., Y_{|{\cal I}|}\}$. We repeat this procedure while updating ${\cal A}$ and ${\cal I}$ and incrementing $i$ until a stopping criteria is met. 
A summary is described in Algorithm \ref{greedy algorithm}.\\
\begin{algorithm}[t]
\caption{USU algorithm}
\label{USU algorithm}
 \KwData{System parameters defined in Section \ref{System Model}}
 \KwResult{Solution for vectors $\boldsymbol{P}^{*}, \boldsymbol{w}^{*}$}
 \vspace{-0.1cm}
 \hbox to \hsize{\dashfill\hfil}
\vspace{-0.1cm}
initialization\;  
$i=1$, $S_1=\{\}$, $T_1=0$, $\boldsymbol{P}^{*}=\boldsymbol{0}$, $\boldsymbol{w}^{*}=\boldsymbol{0}$,\\

\While{$1$}{

1: $P_k=P_{tot}/K,\ \forall k$. \\ 
2: Having $\boldsymbol{P}$ and $\boldsymbol{t}$, use simplex method to efficiently solve \eqref{Boolean relaxed sensor selection}.\\
%
%
3: $S_{i+1}$ is the set of indices corresponding to $i$ largest values of $\boldsymbol{w}$.\\
4: $P_k=P_{tot}/i,\ k\in S_{i+1}$. \\
5: $T_{i+1}=\text{tr}\left(\boldsymbol{J}\left(\{P_k\}_{k\in S_{i+1}}\right)\right)$.\\
\If{$T_{i+1}\leq T_{i}$ ~$\vee$ ~$i\geq K$,}
{$w_k^{*}=1,\ k\in S_{i}$,\\
$P_k^{*}=P_{tot}/i,\ k\in S_{i}$.\\
Return $T_i$ as maximum value of objective function.\\
break
}
$i=i+1$.
}
\vspace{-0.1cm}
\end{algorithm}
\begin{algorithm}[t]
\caption{greedy algorithm}
\label{greedy algorithm}
 \KwData{System parameters defined in Section \ref{System Model}}
  \KwResult{Solution for vectors $\boldsymbol{P}^{*}, \boldsymbol{w}^{*}$}
  \vspace{-0.1cm}
  \hbox to \hsize{\dashfill\hfil}
 \vspace{-0.1cm}
 initialization\;  
 $i\!=\!1$, ${\cal A}\!=\!\{\}$, ${\cal I}\!=\!\{1, \dots, K\}$, $T_1\!=\!0^{+}$, $\boldsymbol{P}^{*}\!=\!\boldsymbol{0}$, $\boldsymbol{w}^{*}\!=\!\boldsymbol{0}$,\\
 
 \While{$1$}{
 
 1: $\boldsymbol{Y}\!=\!\boldsymbol{0}$.\\
 2: \For{$j=1, ..., |{\cal I}|$}
 {2-1: ${\cal U}_j\!=\!{\cal A}\cup {\cal I}_j$.\\
 %
 %
 2-2: Solve \eqref{max tr(J) for k in U_j} to obtain the optimal solution $\{P^{'}_k\}_{k\in {\cal U}_j}$.\\
 2-3: $Y_{j}=\text{tr}\left(\boldsymbol{J}\left(\{P^{'}_k\}_{k\in {\cal U}_j}\right)\right)$.\\
 }
 3: $T_{i+1}=max\ \boldsymbol{Y}$.\\
 \If{$\frac{T_{i+1}-T_{i}}{T_{i}}\leq\epsilon_0$, (i.e., solution converged) ~$\vee$ ~$i\geq K$,}
 {$w_k^{*}=1,\ k\in {\cal A}$,\\
 $\{P_k^{*}\}_{k\in {\cal A}}=\{P^{'}_k\}_{k\in {\cal A}}$.\\
 Return $T_i$ as maximum value of objective function.\\
 break
 }
 4: Update ${\cal A}$ by setting ${\cal A}={\cal U}_{\{\underset{j} {\mbox{argmax}}\ \boldsymbol{Y}\}}$.\\
 5: Remove ${\cal I}_{\{\underset{j} {\mbox{argmax}}\ \boldsymbol{Y}\}}$ from ${\cal I}$.\\
 $i=i+1$.
 }
 \vspace{-0.1cm}
 \end{algorithm}
$\bullet$ \underline{MCKP Algorithm}:
%
%
We discretize $P_{tot}$ into $N$ samples $\boldsymbol{P}_d\!=\![P_{d_1},..., P_{d_N}]^T$ and obtain matrix $\boldsymbol{T}$ such that its $(k,j)$th entry $[\boldsymbol{T}]_{k,j}\!=\!t_k(P_{d_j}),\ k\!=\!1, \dots, K,\ j\!=\!1, \dots, N$. Let $\boldsymbol{W}$ be a $K\!\times\!N$ matrix whose entries can be zeros or ones. Then \eqref{power aloc sen selection problem} becomes:
\begin{align} \tag{P4}\label{MCKP}
\mathop{\text{maximize}}_{\boldsymbol{W}}\ \ \ \ &\boldsymbol{1}^T\underbrace{(\boldsymbol{W}\circ\boldsymbol{T})}_{\mathclap{\text{Hadamard product}}}\boldsymbol{1}\nonumber\\
\text{s.t.}\ \ \ \ &\boldsymbol{1}^T(\boldsymbol{W}\circ(\boldsymbol{1}\boldsymbol{P}_d^T))\boldsymbol{1}\leq P_{tot},\nonumber\\
&[\boldsymbol{W}]_k\boldsymbol{1}=1,\ k\!=\!1, \dots, K,\nonumber\\ 
&\boldsymbol{W}\in \{0,1\}^{K\!\times\!N},\nonumber
\end{align}
in which $[\boldsymbol{W}]_k$ is $k$-th row of $\boldsymbol{W}$. The second constraint states that only one value from vector $\boldsymbol{P}_d$ should be chosen for each sensor, which in fact is the associated power of that sensor. Problem \eqref{MCKP} is a MCKP that is NP-hard, but it can be solved in pseudo-polynomial time through dynamic programming \cite{PISINGER_MCKP}. 
\section{Numerical and Simulation Results} \label{simulation}
We assume a zero mean Gaussian vector
$\boldsymbol{\theta}=\left[\theta_1,\theta_2\right]^T$ with $\boldsymbol{\cal C}_{\boldsymbol{\theta}}=[4,0.5;0.5,0.25]$, and $\sigma_{n_k}\!=\!1$, $\sigma_{\nu_k}\!=\!1$, $|h_k|\!=\!0.7$, $L_k=3$ for all $k$. We choose a proper $\tau_k$ based on the observation model and joint pdf of the unknown vector such that $p(|x_k|\geq\tau_k)\approx0$. To this end, we choose $\tau_k=3\sqrt{\sigma_{n_k}^2+\mathbf{a}_k^T\boldsymbol{\mathcal C}_{\boldsymbol{\theta}}\mathbf{a}_k}$. 

\begin{figure}[h]
	\centering
	\subcaptionbox{\label{homog_max_tr_J}}{\vspace{-.25 cm}\includegraphics[width=3.5in]{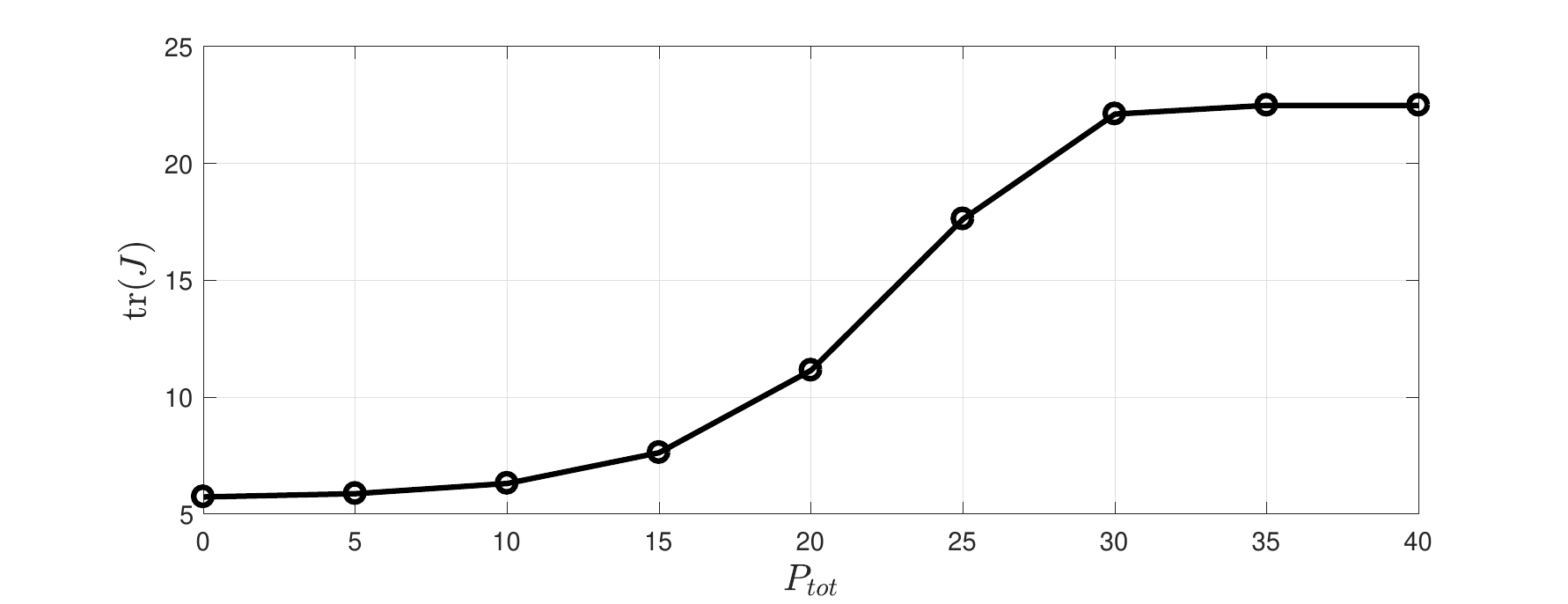}}
	\subcaptionbox{\label{homog_num_of_sen}}{\vspace{-.25 cm}\includegraphics[width=3.5in]{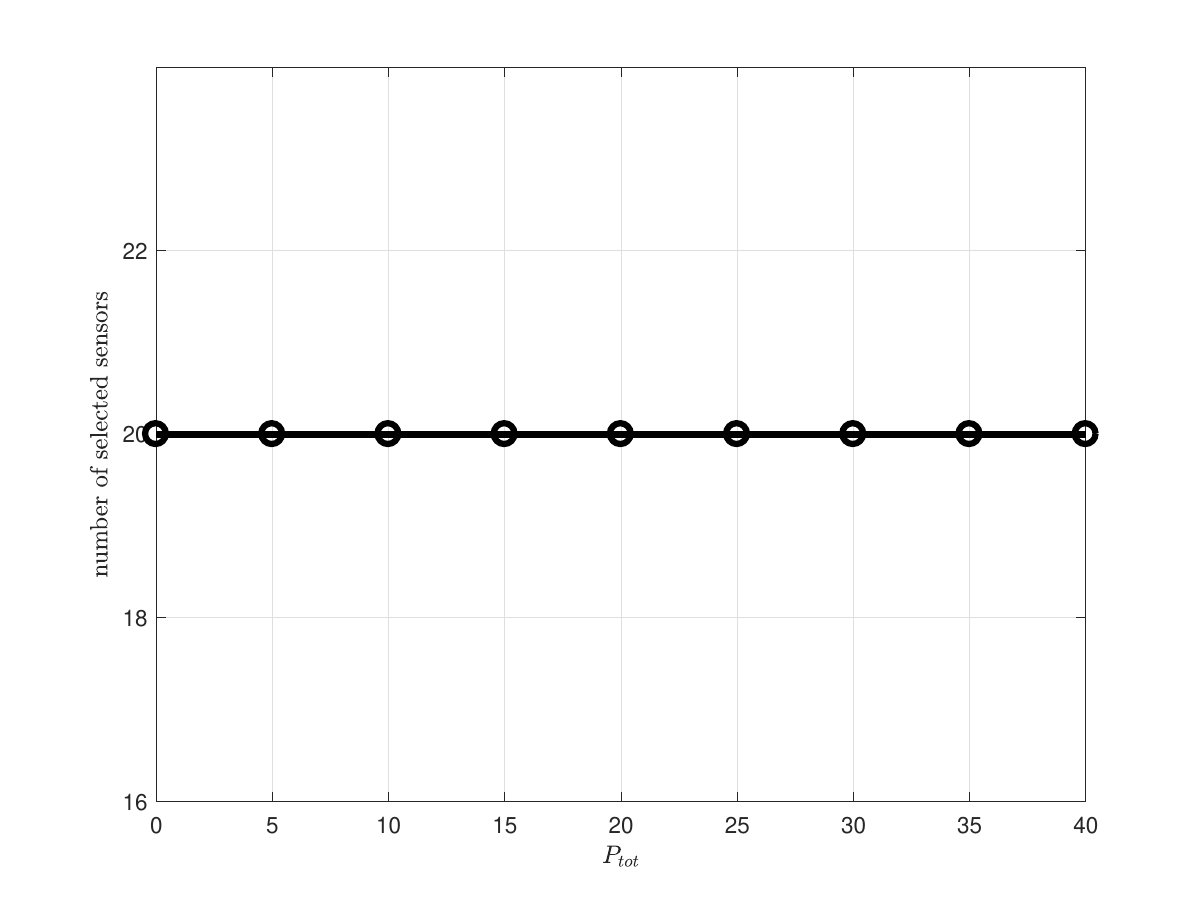}}
	\caption{Homogeneous sensor network: (a) tr($\boldsymbol{J}$), and (b) number of selected sensors, versus $P_{tot}$.}  
	\label{homog} 
\end{figure}

First, we consider a homogeneous sensor network with $\mathbf{a}_k=[0.6,0.8]^T$ for all $k$. Fig.~\ref{homog_max_tr_J} shows that an increase in $P_{tot}$ results in increase in $\text{tr}(\boldsymbol{J})$. Moreover, we are interested to know the optimal number of sensors that maximizes $\text{tr}(\boldsymbol{J})$. From Fig.~\ref{homog_num_of_sen} we observe that for a given $P_{tot}$, it is always beneficial to uniformly allocate $P_{tot}$ among all of the sensors. 

\begin{figure}[h]
	\centering
	\subcaptionbox{\label{heterog_max_tr_J}}{\vspace{-.25 cm}\includegraphics[ width=3.5in]{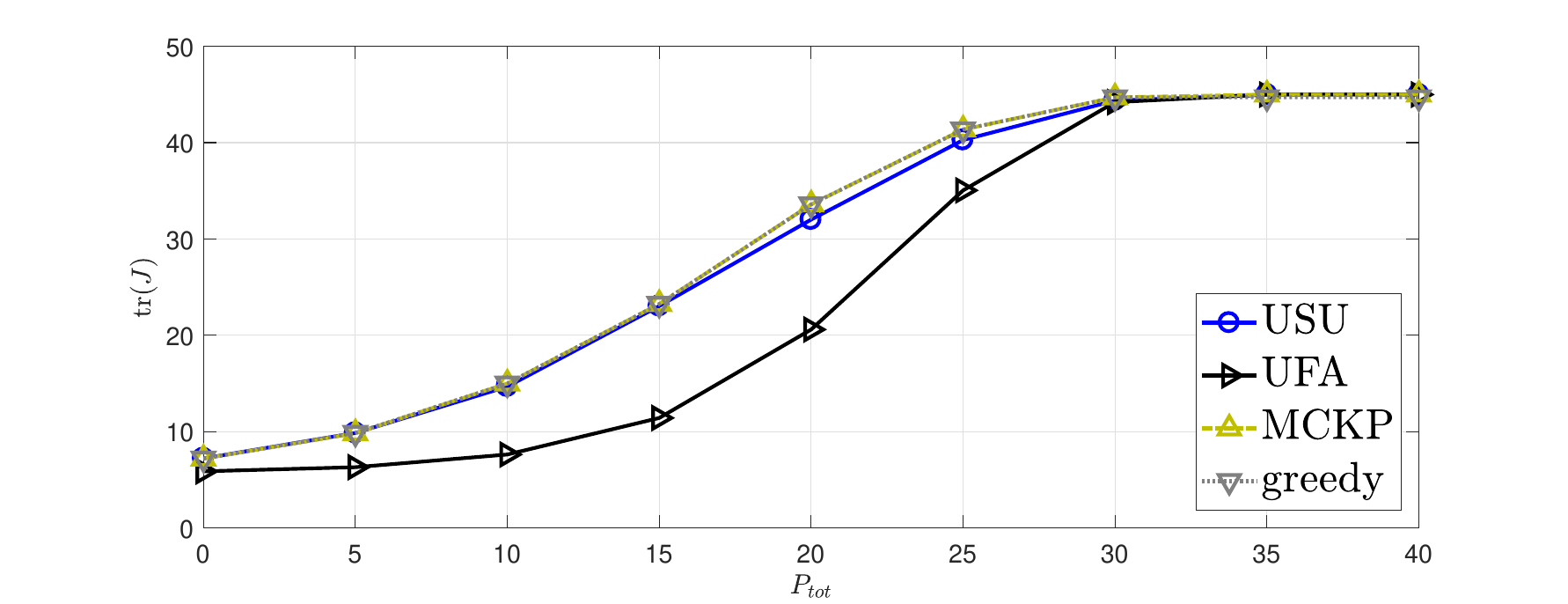}}
	\subcaptionbox{\label{heterog_num_of_sen}}{\vspace{-.25 cm}\includegraphics[width=3.5in]{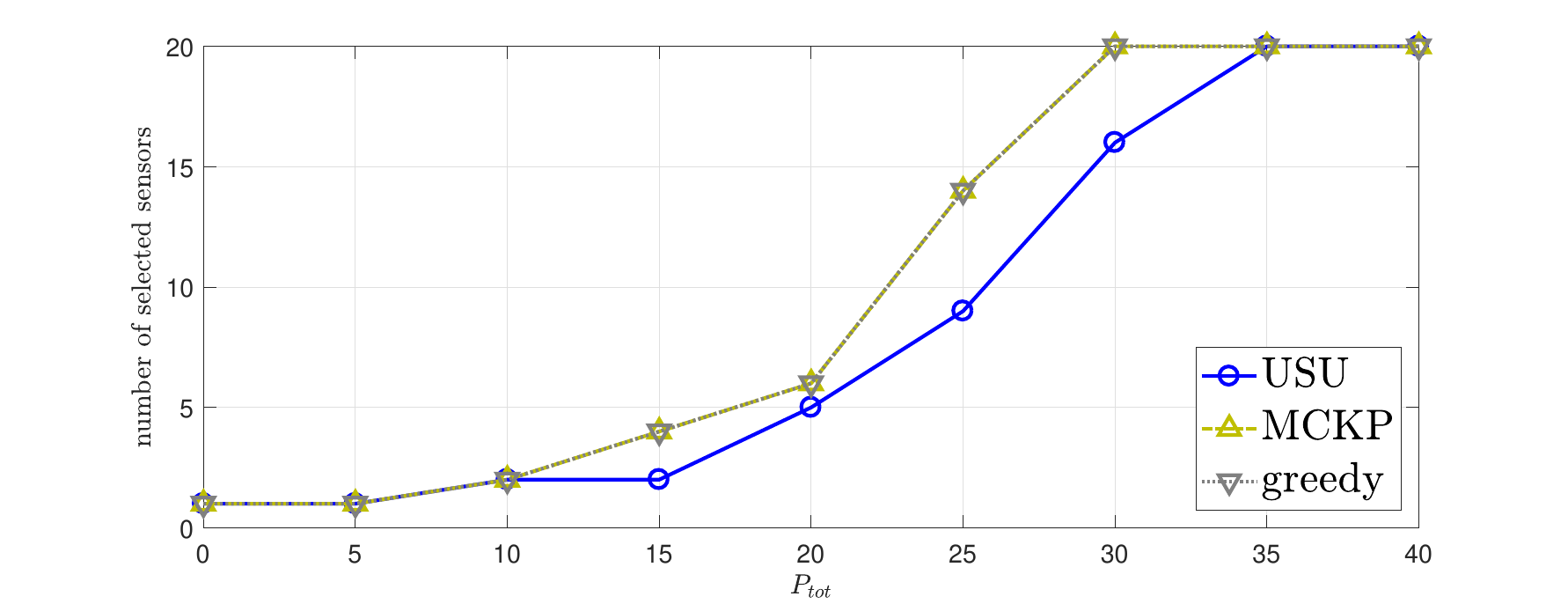}}
	\caption{Heterogeneous sensor network: comparison of (a) tr($\boldsymbol{J}$), and (b) number of selected sensors, versus $P_{tot}$ for different algorithms.}  
	\label{heterog} 
\end{figure}

Next, we consider $K\!=\!20$ sensors that are randomly deployed in a $2m\times2m$ field. Assuming Cartesian coordinate system with origin at center of the field, the goal is to estimate two signal sources $\boldsymbol{\theta}=\left[\theta_1,\theta_2\right]^T$. The distance between $\theta_i$ located at $(x_{t_i},y_{t_i})$ and sensor $k$ located at $(x_{s_k},y_{s_k})$ is: 
\begin{equation*}
d_{ki}=\sqrt{(x_{s_k}-x_{t_i})^2+(y_{s_k}-y_{t_i})^2}, \ \ \ k=1,..., 20, \ \ i=1,2
\end{equation*}
Let $d_{0i}$ be distance of $\theta_i$ from origin, $d_{01}=d_{02}=1m$, and 
%
%
$\mathbf{a}_k^T=[(\frac{d_{01}}{d_{k1}})^n, (\frac{d_{02}}{d_{k2}})^n]$ where $n$ is signal decay exponent which is approximately 2 for distances $\leq\!1km$ \cite{Li_Eurasip_2003}. Fig.~\ref{heterog} compares the performance of 3 algorithms USU, greedy, and MCKP. We also consider uniform power allocation among all sensors, i.e., $\{P_k=P_{tot}/K\}_{k=1}^K$ where no sensor selection is carried out. We refer to this algorithm as UFA algorithm. For MCKP, our simulations suggest that $N=100$ samples are sufficient to reach the optimal solution. Fig.~\ref{heterog_max_tr_J} illustrates that greedy and MCKP algorithms perform similarly, and they both outperform USU algorithm, which is intuitive. However, the performance of USU algorithm is not far from other algorithms, which highlights the advantage of sensor selection even when power is uniformly allocated among sensors. Moreover, all three algorithms significantly outperform UFA algorithm. Fig.~\ref{heterog_num_of_sen} reveals that USU algorithm always selects less number of sensors compared to other algorithms, which can save energy in battery-powered sensor networks. 
%
%
\section{Conclusions} \label{conclusions}
We considered the problem of sensor selection and power allocation to maximize the trace of Bayesian Fisher information matrix $\boldsymbol{J}$ under network transmit power constraint for distributed estimation of a zero mean Gaussian vector in wireless sensor networks. Three algorithms named as USU, greedy, and MCKP algorithms were proposed and their performances were compared numerically. Simulation results demonstarted the superiority of these algorithms compared to the algorithm that allocates power equally among all sensors. Moreover, the advantage of USU algorithm is that its performance is close to those of other algorithms and also, less number of sensors get activated compared to other algorithms, which contributes to energy saving in battery-powered sensor networks.  
\section*{Acknowledgment} \label{acknowledgment} 
This research is supported by NSF under grants CCF-1341966 and CCF-1319770.
\bibliographystyle{IEEETran}
\bibliography{myref}
%
%

%




\end{document}